\documentclass[twocolumn,showpacs,preprintnumbers,amsmath,amssymb]{revtex4}

\usepackage{graphicx}
\usepackage{dcolumn}
\usepackage{bm}
\usepackage{color}

\begin{document}

\preprint{APS/123-QED}

\title{Phase locking a clock oscillator to a coherent atomic ensemble}

\author{R.~Kohlhaas$^{1,2}$}
\author{A. Bertoldi$^3$}
 \email{andrea.bertoldi@institutoptique.fr}
\author{E.~Cantin$^{3,4}$}
\author{A. Aspect$^1$}
\author{A. Landragin$^2$}
\author{P. Bouyer$^{1,3}$}

\address{	$^1$Laboratoire Charles Fabry, Institut d'Optique, CNRS, Universit{\'e} Paris-Sud,
			avenue Augustine Fresnel, F-91127 Palaiseau, France \\
		$^2$LNE-SYRTE, Observatoire de Paris, CNRS and UPMC,
			61 avenue de l'Observatoire, F-75014 Paris, France \\
		$^3$Laboratoire Photonique, Num{\'e}rique et Nanosciences - LP2N Universit{\'e} Bordeaux - IOGS - CNRS: UMR 5298,
			rue Mitterrand, F-33400 Talence, France \\
		$^4$Quantel, 4 rue Louis de Broglie, Building D, F-22300 Lannion, France
	}

\date{\today}

\begin{abstract}

The sensitivity of an atomic interferometer increases when the phase evolution of its quantum superposition state is measured over a longer interrogation interval. In practice,
a limit is set by the measurement process, which returns not the phase, but its projection in terms of population difference on two energetic levels. The phase interval over
which the relation can be inverted is thus limited to the interval $[-\pi/2,\pi/2]$; going beyond it introduces an ambiguity in the read out, hence a sensitivity loss. Here, we
extend the unambiguous interval to probe the phase evolution of an atomic ensemble using coherence preserving measurements and phase corrections, and demonstrate the phase lock
of the clock oscillator to an atomic superposition state. We propose a protocol based on the phase lock to improve atomic clocks limited by local oscillator noise, and foresee
the application to other atomic interferometers such as inertial sensors.

\end{abstract}


\maketitle

\newpage

From the first observations of Huygens on the coordinated motion of coupled nonlinear oscillators \cite{huygens1893}, phase synchronization has evolved to an indispensable tool
for time and frequency metrology and starts to be investigated for quantum systems \cite{mari2013,xu2014,cox2014}. Phase lock loops (PLLs) \cite{guanChyun1996}, where a local
oscillator (LO) is phase locked to a reference signal, are widely used for the generation of atomic time scales \cite{jones2000,diddams2000}, the synchronization in
telecommunication \cite{slavik2010} or in radio navigation \cite{braasch1999}. In usual atomic frequency standards, however, only the frequency of the local oscillator is
locked on the atomic resonance. This feature derives from the quantum nature of the reference system, \textit{i.e.}, the quantum superposition of two internal states of an
ensemble of atoms, molecules or ions, which is destroyed by the detection at the end of the interrogation process. A similar limitation exists for any measurement of a quantum
system as in magnetometers or inertial sensors \cite{arndt2014}. Locking the phase of the local oscillator onto the phase of the quantum superposition of the two levels of the
quantum system would improve the long term stability, as the phase is the integral of the frequency, and reduce the constraints on the stability of the LO or of the measured
signal. More in general, phase locking a classical system to a quantum system would give a direct link in metrology to the fundamental oscillations of quantum particles and
could lead to enhanced sensitivities and new applications in precision measurements. Here, we demonstrate the direct phase lock of a LO to an atomic ensemble, based on repeated
coherence preserving measurements of the atomic ensemble. We also study how this technology could improve atomic clocks subject to local oscillator noise.

In an atomic clock, the frequency of a LO is repeatedly referenced to an atomic transition frequency by comparing their respective phase evolutions in an interrogation time T and
applying a feedback correction. During the interrogation, the atoms are in a superposition state, and the projection of the relative phase between the LO and the atomic ensemble
is measured as a population imbalance of the two clock levels. The readout is thus a sinusoidal function of the phase drift, and the latter can only be unambiguously determined
if it stays within the $[-\pi/2,\pi/2]$ interval, hereafter called inversion region. Hence, for a given LO noise, the interrogation time of the atomic transition must be kept
short enough such that phase drifts beyond the inversion region are avoided. Currently, LO noise limits the interrogation time in ion \cite{chou2010} and optical lattice clocks
\cite{hinkley2013,bloom2014,ushijima2015,nicholson2014} and is expected to become a limit for microwave clocks with the recently discovered spin-self rephasing effect
\cite{kleineBuning2011}. The standard approach to tackle this issue consists in improving the quality of local oscillators
\cite{jiang2011,kessler2012,thorpe2011,cole2013,amairi2013}. As an alternative, it has been recently proposed to track and stabilize the LO phase evolution using several atomic
ensembles probed with increasing interrogation time \cite{rosenband2013,kessler2014}, or by enhancing the Ramsey interrogation interval by stabilizing the LO either via cascaded
frequency corrections \cite{borregaard2013b} or by coherence preserving measurements on the same atomic ensemble and feedback \cite{shiga2012,shiga2014}.

In this letter we show for the first time how the phase lock of a classical oscillator to an atomic superposition state can be exploited to keep the relative phase between the LO
and the atomic system in the inversion region. In addition, we demonstrate a protocol based on this phase lock to operate an atomic clock beyond the limit set by the LO
decoherence, which nowadays represents the bottleneck of the best available frequency standards. We begin with a minimally destructive measurement of the LO phase drift when it
is within the inversion region; the measurement readout is then used to correct the LO phase so as to reduce its drift. The cycle is repeated using the residual atomic coherence.
The result is a series of successive, phase-related measurements of the relative phase evolution, and the feedback keeps the phase in the inversion region, leading to an
effectively longer interrogation time \cite{note1}. We demonstrate this approach with a trapped ensemble of neutral atoms probed on a microwave transition.

\begin{figure}[!h]
\centering
\includegraphics[width=8.6cm]{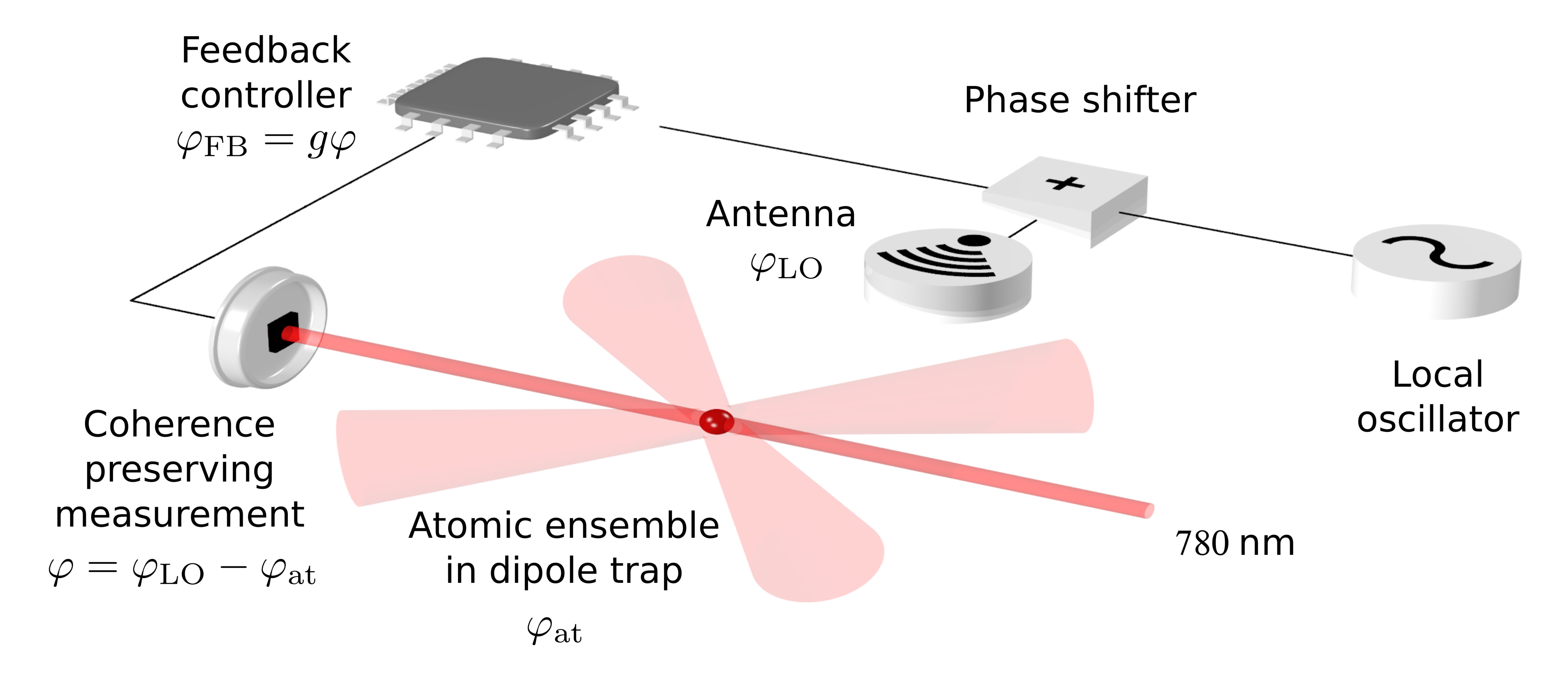}
\caption{\label{fig:pic1} \textbf{Experimental scheme.} The evolution of the LO phase $\varphi_{\rm{LO}}$ is compared to the phase $\varphi_{\rm{at}}$ of an atomic ensemble in a
superposition state using coherence preserving measurements in a Ramsey spectroscopy sequence. The relative phase is obtained from the read out of the population difference, and
is used to implement the phase lock between the two oscillators by applying a feedback correction phase $\varphi_{\rm{FB}}$ on the LO output using a phase actuator. The light
shift induced by the optical trap and by the probe has been engineered to have a homogeneous measurement of the atomic ensemble \cite{vanderbruggen2013}.}
\end{figure}

\begin{figure}[!h]
\centering
\includegraphics[width=8.6cm]{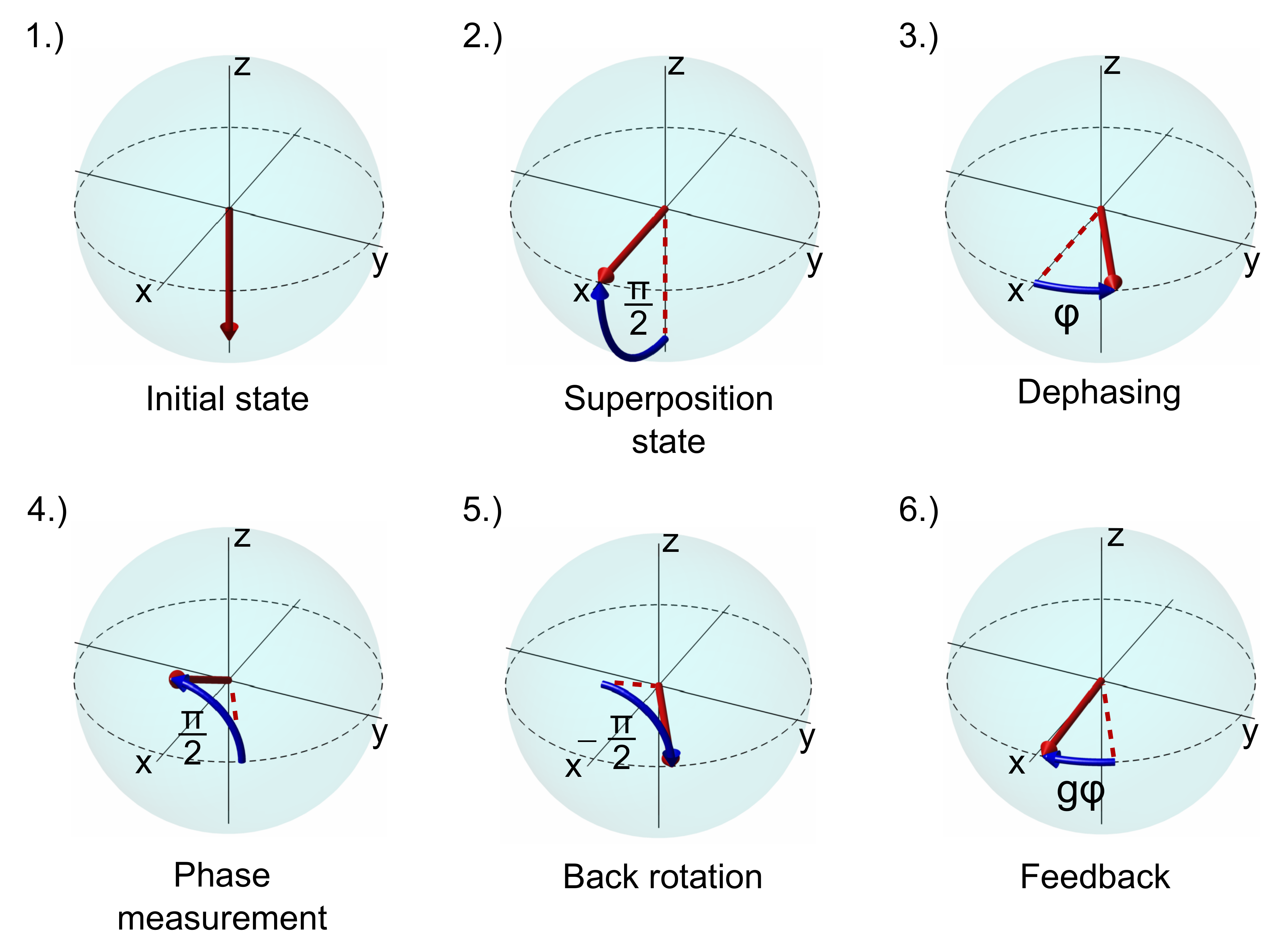}
\caption{\label{fig:pic2} \textbf{Bloch's sphere representation of the phase lock between the LO and the atomic superposition state.} The phase lock between the classical
oscillator and the atomic spin is obtained using repeated, time correlated Ramsey interrogations and feedback. The sequence begins by preparing the atomic CSS in the  $\left|
\, \downarrow \, \right\rangle$ state via optical pumping (step 1.). The measurement of the relative phase between the CSS and the LO starts when a $\pi$/2 microwave pulse
around the $y$ axis brings the CSS into a balanced superposition of the $\left| \, \downarrow \, \right\rangle$ and $\left| \, \uparrow \, \right\rangle$ states, depicted as
a vector on the equatorial plane of the Bloch sphere (step 2.). The $x$ axis is chosen to represent the phase of the local oscillator, and $\varphi$ the relative phase
between the LO and the atomic superposition that evolves because of the LO noise (step 3.). After an interrogation time T, the projection of $\varphi$ is mapped onto a
population difference by a projection $\pi$/2 pulse around the $x$ axis and read out with the coherence preserving detection (step 4.). The CSS is rotated back to the
equatorial plane by a reintroduction $\pi$/2 pulse around the $x$ axis (step 5.), and feedback is applied on the phase of the LO (step 6.). The PLL between the LO and the
atomic ensemble consists in the repetition of the steps from 3.) to 6.), potentially till the atomic ensemble shows a residual coherence.}
\end{figure}

The experimental scheme shown in Fig. \ref{fig:pic1} has been described in \cite{vanderbruggen2013}. A cloud of cold $^{87}$Rb atoms is trapped in an optical potential (see
Appendix \ref{appA}), prepared with a $\pi$/2 pulse of a resonant microwave field in a balanced superposition state of two hyperfine levels $\left| \, \downarrow \,
\right\rangle \equiv \left| \, F=1,m_{F}=0 \, \right\rangle$ and $\left| \, \uparrow \, \right\rangle \equiv \left| \, F=2,m_{F}=0 \, \right\rangle$ of the electronic ground
state, and probed using a nondestructive detection. Typically, $5 \times 10^5$ atoms at a temperature of 10 $\mu$K are used in the measurements reported here. The phase
$\varphi_{\rm{at}}$ of the superposition state oscillates at a frequency of 6.835 GHz corresponding to the energy difference between the $\left| \, \downarrow \,
\right\rangle$ and $\left| \, \uparrow \, \right\rangle$ atomic states, which is the fundamental reference if atoms are protected from perturbations. A microwave LO has a
frequency close to the atomic frequency difference, so that the relative phase $\varphi = \varphi_{\rm{LO}} - \varphi_{\rm{at}}$ between the two oscillators drifts slowly
because of the LO noise. $\varphi$ can be measured using the Ramsey spectroscopy method (see Fig. \ref{fig:pic2}): a second $\pi$/2 microwave pulse (projection pulse) maps it
onto a population difference, which we read out with a weak optical probe perturbing the atomic quantum state only negligibly and preserving the ensemble coherence
\cite{smith2006,lloyd2000,vanderbruggen2013}. Unlike for destructive measurements, the interrogation of $\varphi$ can continue in a correlated way, once the action of the
projection pulse is inverted using an opposite $\pi$/2 microwave pulse (reintroduction pulse), which brings the atomic state back to the previous coherent superposition.
Moreover, after each measurement and reintroduction pulse, the phase read out can be used to correct the LO phase. The evolution and manipulation of the atomic ensemble can
be illustrated using the Bloch sphere representation (Fig. \ref{fig:pic2}): the collective state of $N_{\rm{at}}$ two-level atoms in the same pure single particle state (also
called coherent spin state (CSS)) forms a pseudo-spin with length J = $N_{\rm{at}}/2$, where $J_z$ denotes the population difference and $\varphi = \arcsin{\left(J_y /
J_x\right)}$ is the phase difference between the phase of the LO and that of the superposition state. A resonant microwave pulse determines the rotation of \textbf{J} around
an axis in the equatorial plane of the Bloch sphere, and the axis direction is set by the phase of the microwave signal. The repetition of the manipulation, measurement and
feedback cycle implements the phase lock of the LO on the atomic superposition state, as shown in steps 2.-6. of Fig. \ref{fig:pic2}.

\begin{figure}[!h]
\centering
\includegraphics[width=8.6cm,keepaspectratio]{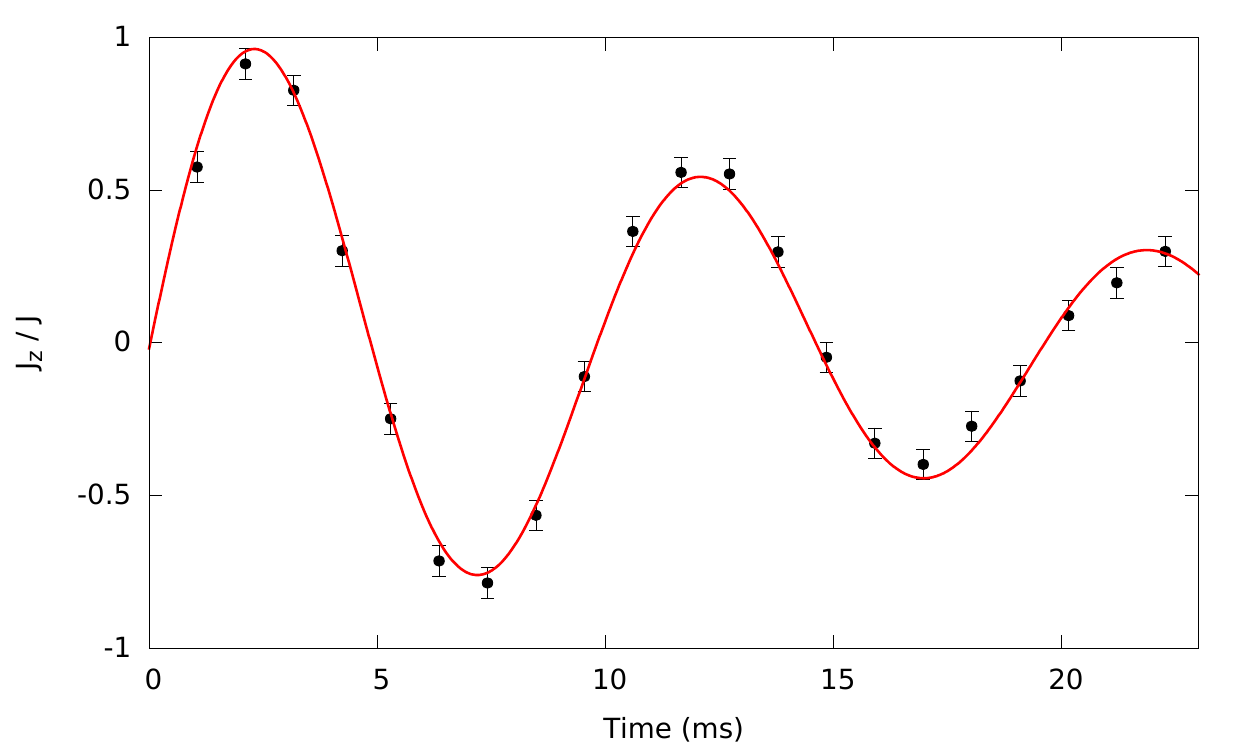}
\caption{\label{fig:pic3} \textbf{Coherence preserving measurement of the relative phase between the LO and the atomic superposition.} Real-time measurement of the normalized population difference to which the relative phase $\varphi$ between the local oscillator and the atomic superposition state is periodically mapped via microwave rotation pulses.
The phase precession is induced by setting the LO frequency 100 Hz off the nominal atomic transition frequency. The experimental points are fitted with a sinusoidal
evolution, damped because of the decoherence induced by the $J_z$ measurement and the residual differential light shift on the clock transition.}
\end{figure}

We first show that we can reconstruct the time evolution of the relative phase between the LO and the CSS by monitoring the population difference and without applying feedback.
For this purpose, we frequency offset the LO by 100 Hz from the nominal resonance, and periodically measure the projection of the relative phase $\sin ( \varphi )$ with only a
small reduction of the atomic ensemble coherence (Fig. \ref{fig:pic3}). Every 1 ms $\varphi$ is mapped to a population difference via a projection $\pi / 2$ microwave pulse
around the x-axis, a weak measurement of $J_z$ is performed, and the collective spin is rotated back to the equatorial plane of the Bloch sphere via a reintroduction $\pi / 2$
pulse around the x-axis. The $\pi / 2$ microwave pulses are derived from an amplified version of the LO at 6.835 GHz leading to a pulse length of $\tau_{\pi / 2}$=47 $\mu$s, and
the rotation axis is controlled with a quadrature phase shifter. The coherence preserving, dispersive measurement relies on frequency modulation spectroscopy (see Appendix
\ref{appB}). Fig. \ref{fig:pic3} shows in a single experimental run how the spin state evolves around the equator of the Bloch sphere, with the relative phase $\varphi$ mapped on
the normalized population difference $J_z/J$. The signal-to-noise-ratio (SNR) of the weak measurements is 20 for a full state coherence and each readout of the relative phase
drift reduces the state coherence by 2\%. The destructivity from the probe is the main decoherence source till 10 ms, then the inhomogeneous light shift of the dipole trap on the
clock states becomes the dominant decoherence source.

\begin{figure}[!h]
\centering
\includegraphics[width=8.6cm,keepaspectratio]{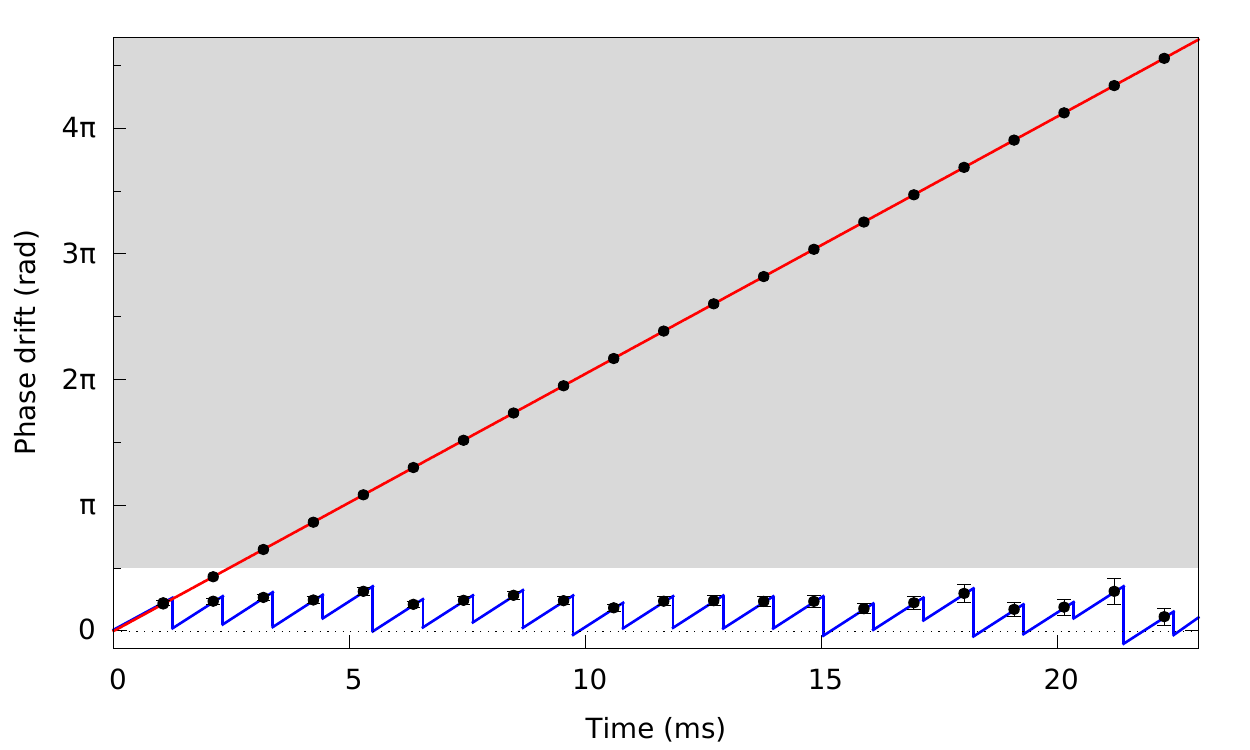}
\caption{\label{fig:pic4} \textbf{Phase lock between the LO and the atomic superposition state.} The evolution of the LO-atom relative phase is reconstructed from the $J_z$
signal in Fig. \ref{fig:pic3} (red, solid line) and when feedback is applied on the phase of the LO after each measurement (blue, solid line). In the open loop case, the
points are reported at the values given by the fit in the previous figure. In the closed loop case, the relative phase is always smaller than $\pi/2$, and does not enter in
the region represented in grey where it cannot be univocally determined from the measurement.}
\end{figure}

\begin{figure}[!h]
\centering
\includegraphics[width=8.6cm,keepaspectratio]{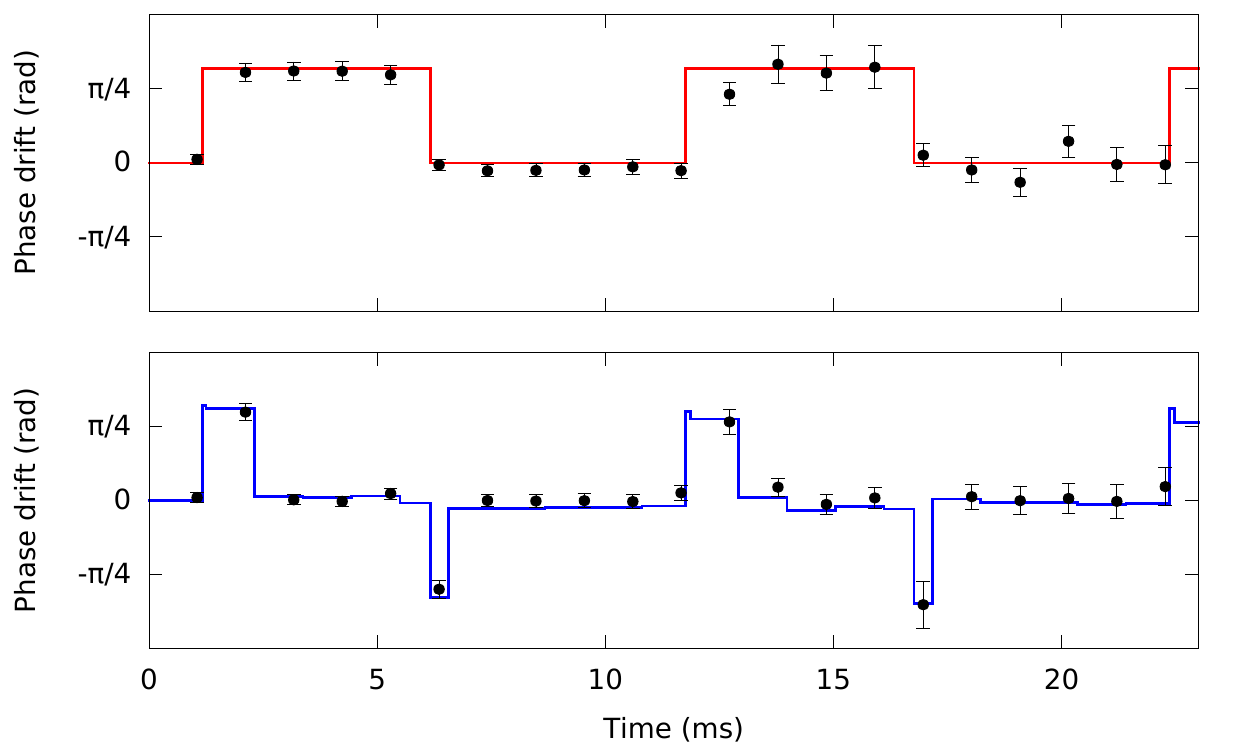}
\caption{\label{fig:pic5} \textbf{Correction of phase jumps between the LO and the atomic ensemble.} Evolution of the LO-CSS phase when periodic phase jumps of $\pi$/3 are
applied back and forth on the LO: the points are obtained from the measured population difference, whereas the solid line represents the LO phase measured after the
correction phase actuator. The controller, implemented using coherence preserving measurements and feedback on a phase actuator, maintains the relative phase close to
zero. Above the system is operated in open loop, below in closed loop. In the latter case, the phase corrections are applied 150 $\mu$s after each measurement.}
\end{figure}

We next introduce feedback and demonstrate that we can phase lock the LO on the atomic superposition state, and increase the Ramsey interrogation time beyond the limit set by the
inversion region between $J_z$ and the relative phase. We apply on the local oscillator two types of signals, first a frequency offset, and second periodic phase jumps, and use
the output of the coherence preserving measurements to actively minimize $\varphi$. The phase lock is obtained by controlling the phase of the local oscillator by means of a
digital phase shifter (see Appendix \ref{appD}). The feedback is performed after the atomic spin is rotated back to the equatorial plane of the Bloch sphere. When the disturbance
applied on the local oscillator consists of a frequency offset, there is a linear phase drift between the LO and the atomic phase (Fig \ref{fig:pic4}, red). The phase evolution
in open loop is reconstructed from the data of Fig. \ref{fig:pic3} by taking into account the damping on the sinusoidal signal due to the decoherence sources, and knowing that a
constant frequency offset is applied on the LO. We remark that sudden sign inversions of the applied frequency offset when $\varphi=\pm \pi/2$ would produce exactly the same
evolution of the population difference, illustrating the need to keep $\varphi$ in the inversion region. In closed loop, the phase drift due to the 100 Hz frequency offset on the
LO is periodically reset to zero, with a precision set by the $\pi$/32 step size of the digital phase shifter and the uncertainty of the coherence preserving measurements. As a
consequence, the $J_z/J$ signal shows a saw-tooth-like evolution (Fig. \ref{fig:pic4}, blue signal). Without phase lock, the phase drift leaves the inversion region after 2.5 ms
and rotates several times around the Bloch sphere, whereas with phase lock it stays in the inversion region for all the 22 ms interval shown in the image. When the feedback is
active the total phase drift results as the phase measured at the end of the Ramsey interferometer, added to the correction phase shifts on the LO via the feedback controller. We
next apply periodic phase jumps of $\pi$/3 back and forth on the LO using a second phase shifter. The signal obtained in open loop is shown at the top of Fig. \ref{fig:pic5}.
When the feedback controller is active, the jumps detected on the relative phase are corrected to zero, with a precision set by the resolution of the phase shifter and the
uncertainty of the weak measurements (Fig. \ref{fig:pic5}, bottom). The solid lines in Fig. \ref{fig:pic4} and \ref{fig:pic5} are drawn from the known timing for the applied
phase signal and the feedback on the phase. For a combination of a phase drift and phase jumps, the relative phase can leave the inversion region while the noise action cannot be
predicted from previous measurements. This problem affects atomic clocks and highlights the requirement of feedback on the LO phase to keep track of the relative phase drifts.
Without feedback, the Ramsey interrogation time should be kept sufficiently short to avoid ambiguities for the measured phase shift.

\begin{figure}
\centering
\includegraphics[width=8.6cm]{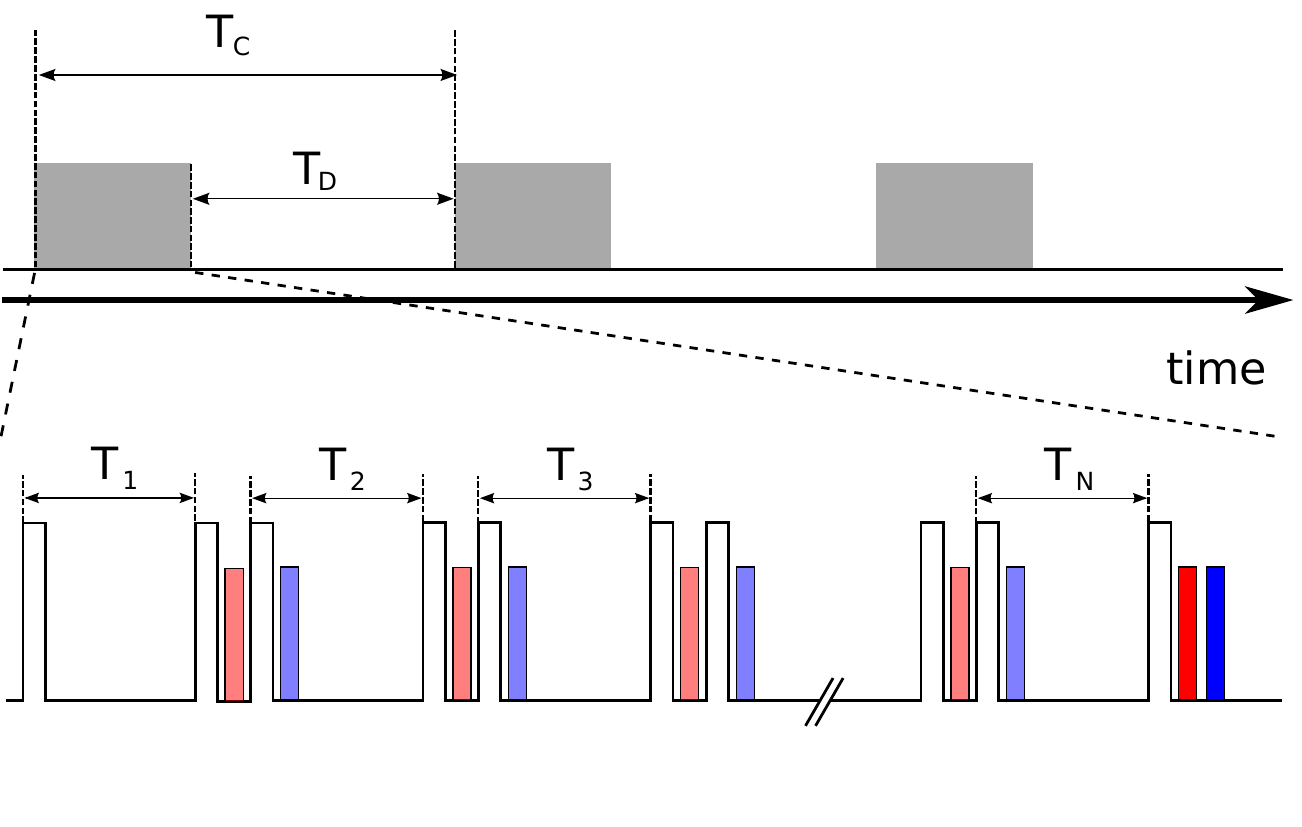}
\caption{\label{fig:pic6} \textbf{Protocol for atomic clock with PLL.} Operation of the clock: at each cycle of duration T$_{\rm{C}}$, the relative phase is repeatedly
measured in a coherence preserving way during the phase lock interval (above: shaded grey areas). Each interrogation, represented in the inset by a light red peak between the
manipulation $\pi/2$ pulses, is followed by a phase correction $\varphi_{\rm{FB}}^{(i)}$ on the LO, represented in the inset by a light blue peak in the right after the
reintroduction of the spin on the equatorial plane of the Bloch sphere. The final phase readout $\varphi_{\rm{f}}$ (dark red peak in the inset), whose SNR is set by the
residual coherence, together with the previously applied phase shifts on the LO, provides the total phase drift $\varphi$ experienced during the extended interrogation
interval T$_{\rm{tot}}$=N$\times$T. The interrogation sequence ends with the application of a frequency correction on the LO (dark blue peak in the inset), then a new atomic
ensemble is prepared in the dead time interval T$_{\rm{D}}$ for the next cycle.}
\end{figure}

We propose now a protocol to efficiently use the phase lock to improve an atomic clock. In a conventional atomic clock, the phase drift $\varphi$ is destructively read out after
a single interrogation and feedback is performed on the LO by the addition of a feedback correction frequency $\omega_{\rm{FB}} = - \varphi / T $ considering unity gain. Our
protocol of using the PLL between the LO and the atomic superposition state in an atomic clock is based on the reconstruction of the phase drift experienced by the LO over the
extended interrogation time T$_{\rm{tot}}$=N$\times$T (N is the number of phase coherent interrogations) by combining the phase shifts applied by feedback and the final phase
readout (Fig. \ref{fig:pic6}). The known phase corrections applied to the LO phase serve the dual purpose of keeping the relative phase in the inversion region and giving a
coarse estimate of the phase drift during T$_{\rm{tot}}$. The final phase measurement $\varphi_{\rm{f}}$, together with the phase corrections $\varphi_{\rm{FB}}^{(i)}$, gives a
precise estimate of the phase drift during T$_{\rm{tot}}$. The total phase drift can be computed as $\varphi_{\rm{tot}} = \varphi_{\rm{f}} -
\sum_{i=1}^N{\varphi_{\rm{FB}}^{(i)}}$, where the sum is over all the correction phase shifts applied by the feedback controller and stored in the microcontroller during the
sequence. The feedback on the frequency is set accordingly to be $\omega_{\rm{FB}}=-\varphi_{\rm{tot}} / \rm{T}_{\rm{tot}}$. After the final phase read out, the phase shifter is
reset to its initial position to avoid any impact of the intermediate phase shifts on the long term clock stability. The measurement SNR of the total phase drift depends only on
the final phase measurement, since it can make use of all the residual ensemble coherence, whereas the SNR of the intermediate measurements has to be only sufficient to keep the
atomic state in the inversion region. To further improve the protocol, the clock operation could be optimized by adapting the destructivity of the intermediate measurements to
the varying coherence of the atomic sample, or implementing an adaptive measurement protocol for the last measurement \cite{borregaard2013a}.

\begin{figure}
\centering
\includegraphics[width=8.6cm]{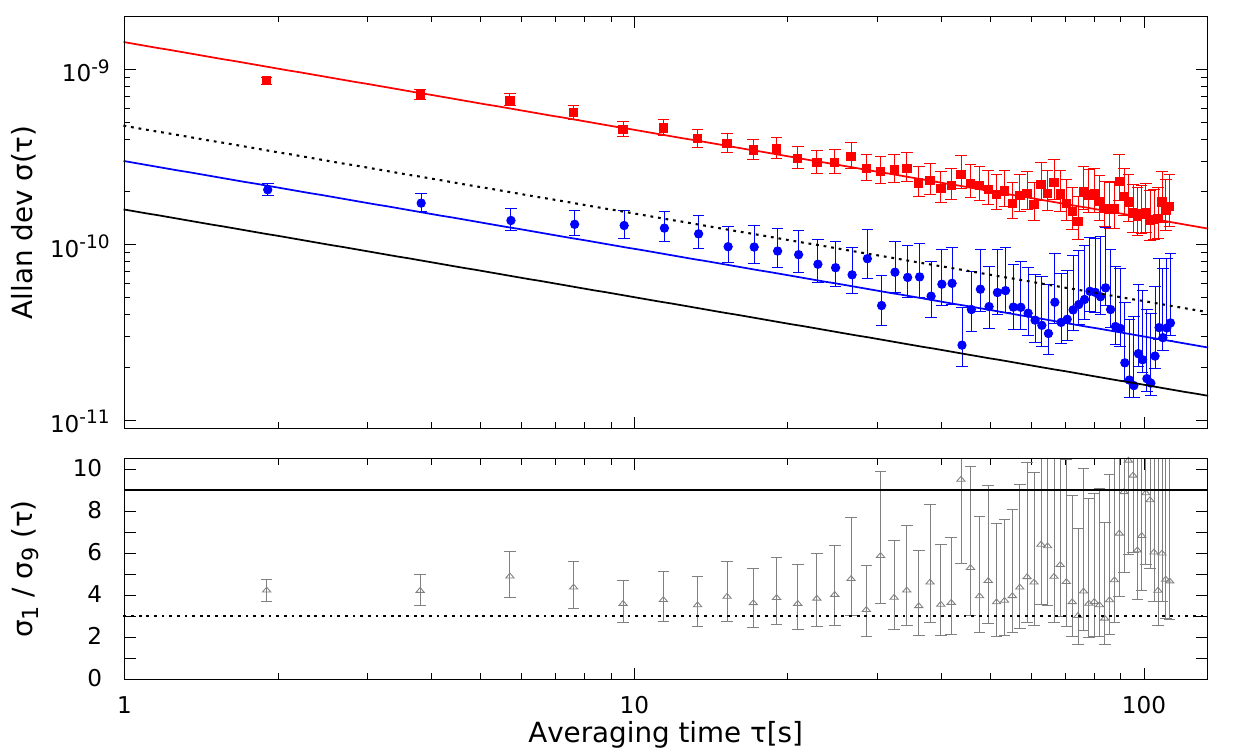}
\caption{\label{fig:pic7} \textbf{Atomic clock implementing a PLL.} (Top) Allan frequency standard deviation $\sigma_1$ for a normal Ramsey clock with interrogation time T=1
ms (red) and $\sigma_9$ for a clock implementing the phase lock between the LO and the atomic superposition state for 9 successive, correlated interrogations on the same
atomic ensemble, for a total interrogation time of 9$\times$T=9 ms (blue). The dead time is in both cases $T_{\rm{D}}$ = 1.9 s. The red and blue lines are fits to the data,
with a slope set to $\tau^{-1/2}$. The continuous black line lies a factor 9 below the red curve, and represents the best achievable level of the phase lock sequence for the
same number of interrogations. The dashed black line lies a factor 3 below the red curve, and is the optimum level for 9 consecutive uncorrelated Ramsey measurements with
duration T each and the same total cycle time. (Bottom) The grey triangles represent the ratio of the Allan deviation for the two clocks; the solid and dashed lines
correspond to the factors 9 and 3 from the top plot.}
\end{figure}

For a proof-of-concept demonstration, we run an atomic clock that exploits the PLL between the LO and the atomic superposition state, and the phase reconstruction protocol. The
LO signal was deteriorated so as to have an increased phase drift over the interrogation interval (see Appendix \ref{appC}). As a benchmark, we first run an atomic clock
adopting a standard Ramsey interrogation sequence, and without any feedback on the phase. In the clock operation,the dead time for the preparation of the new ensemble is
$T_{\rm{D}} = 1.9$ s and the interrogation time is set to T = 1 ms, much shorter than the measured atomic coherence lifetime. The phase measurement is performed with the
coherence preserving detection adopted for the phase lock. The two-sample Allan frequency standard deviation was calculated from the sum of the LO noise and the correction
signal applied by the feedback controller. The clock instability reaches a $\tau^{-1/2}$ scaling after a few clock cycles (see Fig. \ref{fig:pic7}, red) at a level consistent
with an initial SNR = 20 and considering the coherence decay in the optical trap. The instability is far higher than with state-of-the art atomic clocks since the experimental
setup was not explicitly designed for the operation of an atomic clock, and the Allan frequency standard deviation is $1.5 \times 10^{-9}$ at 1 s. We then operate an atomic
clock making use of a PLL sequence with N=9 successive interrogations, and again with T=1 ms, thus increasing the total interrogation time to 9 ms (see Fig. \ref{fig:pic6}).
For simplicity, the intermediate and the final phase readouts are set to have the same measurement strength. The Allan frequency deviation shows a $\tau^{-1/2}$ scaling as
expected for atomic clocks, which demonstrates that the phase reconstruction protocol is working properly. In the opposite case, frequency offsets would be corrected only at a
short time with the phase actuator, and the stability of the clock would then diverge because of the non-zero dead time interval. The comparison of the instability shows that
the clock adopting the phase lock and reconstruction method is at a lower level by a factor (4.76$\pm$0.25) with respect to the clock implementing the standard Ramsey
interrogation, as shown in Fig. \ref{fig:pic7} (blue).  In the optimal case, the instability would decrease by a factor 9 (Fig. 7, black line) because of the correspondingly
longer interrogation time. Experimentally, we obtain a lower value because of several detrimental effects, where the main contributions are a reduced SNR due to the cumulated
destructivity from the probe and the decay from the optical dipole trap, and the finite phase shifter accuracy, equal to 38 mrad. The result can be further compared to the case
that the 9 phase measurements would have been uncorrelated, for example by repreparing the atomic state and starting a new Ramsey cycle after each measurement
\cite{lodewyck2009}. We would expect here a factor 3 (Fig. 7, black dashed line), which is clearly exceeded by the phase lock sequence.

The phase lock can be performed as long as the coherence of the state is maintained. For integration times longer than the coherence lifetime of the trapped ensemble, the atomic
phase is lost and our locking scheme becomes again a frequency lock, like when the quantum superposition is destroyed by the detection. In our experiment, the coherence lifetime
is limited to 20 ms by the dephasing in the optical dipole trap. Nevertheless, trapped induced dephasing of the atomic state can be suppressed for $^{87}$Rb as reported in
\cite{kleineBuning2011,radnaev2010}, whereas in an optical lattice it is strongly reduced with the choice of light at the magic wavelength \cite{katori2003}. In the original
proposal to lock the local oscillator phase on the atomic phase \cite{shiga2012}, frequency feedback on the local oscillator after each weak $J_z$ measurement is performed. This
scheme leads to a longer effective interrogation time, but to a SNR given by the weak measurements, which is lower than that of a measurement at the quantum projection noise and
beyond. Our protocol can overcome this limit, and reach projection limited readout while keeping an extended interrogation time, thanks to the feedback on the LO phase.

In the phase lock sequence, several effects must be considered to maximize the SNR of the last measurement while maintaining a high accuracy on the total phase drift over the
increased interrogation time: the rotations operated on the Bloch sphere must be fast, the measurement induced decoherence limited, and the phase shifter used for the
correction accurate. The decoherence related to the repeated interrogations of the relative phase can be strongly reduced by the use of an optical cavity to enhance the probe
interaction with the atomic ensemble \cite{bohnet2014,lee2014,leroux2010}. An optimized clock configuration would consist in a two atomic ensembles using the same LO: the
first ensemble provides the information to implement the phase feedback algorithm on the LO; the resulting corrected phase for the LO stays in the inversion region for a much
longer period, and this prestabilized LO is used to interrogate the master ensemble with the standard Ramsey sequence. This scheme avoids the requirement of a trade-off
between the number of intermediate measurements and the SNR of the final measurement by separating the two problems. It also removes the systematics imposed by the
intermediate coherent manipulations and measurements of the atomic state, which now affects only the first ensemble. The solution promises the same benefits foreseen for the
phase reconstruction schemes proposed in \cite{rosenband2013,kessler2014}, but using only a single additional ensemble.

In an atomic clock the phase lock between the LO and the atomic superposition state can reduce the Dick effect \cite{dick1987}, \textit{i.e.} the aliasing of the clock
oscillator, thanks to the longer interrogation time. However, the most important advantage of the scheme is the reduction of the decoherence related to the local oscillator,
which translates to a lower white noise frequency for a fixed detection noise. The noise reduction can be exploited to lower the LO stability requirements to the benefit of other
parameters, like portability of the experimental setup, for mobile or spatial applications, or viceversa, to remove the limitation set by the LO to reach ultimate performances. 

For example, the interrogation interval in the best available optical clocks \cite{hinkley2013,bloom2014,ushijima2015,nicholson2014} is limited to $\simeq1$ s by the quality of the local oscillator; this limitation is not fundamental, and our method could increase the interrogation interval and hence the sensitivity in a way at best proportional to the number of coherence preserving interrogations. LOs with higher coherence will reduce the number of intermediate operations required to obtain a given clock interrogation interval. The latter will then be limited by other effects: practically, a first limit is set by the vacuum quality, which can reduce the ensemble coherence via background collisions, but extremely long trapping lifetimes have been already reached for trapped atoms [39]. Ultimately, with the combination of existing techniques and the method presented in this paper, the 1--10 mHz excited states lifetime expected for alkaline earth-like atoms could be reached, which motivates the search for transitions with a lower linewidth \cite{campbell2012,derevianko2012,kozlov2013}.

Phase locking the LO to the atomic state preserves classical correlations in time against the decoherence by the local oscillator. The technique and the related enhancement
factor could thus be combined with spin squeezing, which improves the clock sensitivity by introducing quantum correlations between the particles to go below the standard
quantum limit \cite{leroux2010,louchetChauvet2010}. More generally, increasing the interrogation time using minimally destructive measurements and feedback on the phase could
be applied to other atomic interferometers, such as atomic inertial sensors, where for example the phase of the interrogation lasers could be locked to the phase evolution of
matter waves.

In conclusion, using a coherence preserving detection we tracked the phase evolution of an atomic collective superposition, and we reproduced it on a classical replica by
introducing feedback to implement a PLL. Phase locking a classical oscillator to an atomic superposition state can be a key technology to improve the sensitivity of systems where
the atomic phase evolution is used to improve the quality of the classical counterpart, such as in atomic interferometers or frequency combs \cite{cadarso2014}, whenever the main
decoherence source is determined by the classical subsystem. This development may open new directions and possibilities in several technological fields and as well in basic
science.

\begin{acknowledgments} We acknowledge funding from DGA, CNES, EMRP (JRP-EXL01 QESOCAS), the European Union (EU) (iSENSE), ANR (MINIATOM), LAPHIA (APLL-CLOCK, within
ANR-10-IDEX-03-02) and ESF EuroQUAM. The EMRP is jointly funded by the EMRP participating countries within EURAMET and the European Union. LCFIO and SYRTE are members of the
Institut Francilien de Recherche sur les Atomes Froids (IFRAF). We thank T. Vanderbruggen for his contribution during the initial phase of the experiment, as well as M.
Prevedelli, G. Santarelli and T. Udem for useful discussions. P.B. acknowledges support from a Chair of Excellence of R{\'e}gion Aquitaine, E.C. from Quantel. Any mention of
commercial products does not constitute an endorsement. \end{acknowledgments}

\appendix

\section{\label{appA}Atomic sample preparation}

$^{87}$Rb atoms laser cooled with a magneto-optical trap are transferred to an optical dipole trap at 1560 nm, that uses a 4 mirror optical resonator to enhance the laser
intensity \cite{bernon2011}. The atoms are trapped at the crossing of two cavity arms with a waist of 100 $\mu$m. The ensemble is evaporatively cooled by decreasing the
intensity of the dipole trap till a temperature of 10 $\mu$K is reached for 5$\times$10$^5$ atoms in a cloud with 1/$e^2$ radius of 50 $\mu$m. In the last operation before
starting the Ramsey interrogation the atoms are optically pumped in the $\left| \, \downarrow \, \right\rangle$ state. The sequence to prepare the ensemble in the initial
state, which corresponds to the dead time $T_{\rm{D}}$ in the atomic clock sequence, lasts 1.9 s.

\section{\label{appB}Non-destructive dispersive probe}

The measurement of $J_z$ is based on the dispersion caused by the trapped atoms on a far off-resonance optical probe. The probe beam has a waist of 47 $\mu$m matched to the size
of the atomic cloud. It is phase modulated at a frequency of 3.853 GHz and frequency referenced at 3.377 GHz on the red of the the F=1$\rightarrow$F'=2 transition; these
conditions produce a symmetric mixing of the $\left| \, \downarrow \, \right\rangle$ and $\left| \, \uparrow \, \right\rangle$ states because of probe induced spontaneous
emission, thus avoiding a vertical offset when the Bloch sphere contracts. In this way, each sideband mainly probes the population of one of the two levels, with the same
magnitude and opposite sign for the couplings. We cancel the probe induced light shift and the related decoherence by precisely compensating the effect of the carrier with that
of the sidebands; the cancellation is obtained by setting a modulation depth of $14.8\%$ for the phase modulation. The differential light-shift on the D2 line from the dipole
trap at 1560 nm was compensated with light blue detuned from the $5^2P_{3/2}\rightarrow 4^2D_{5/2,3/2}$ transitions at 1529 nm. When the total power of the probe is set to 480
$\mu$W, it causes the decay of the atomic coherence with a lifetime of 2.85 $\mu$s; in the experiment here reported the interrogation pulses, obtained using an amplitude
electro-optic modulator, have been set to last 60 ns. Each pulse determines then a 2\% destructivity of the ensemble coherence, and a SNR of 20 for the $J_z$ measurement on
the initial sample of 5$\times$10$^5$ atoms. The population imbalance read-outs have been normalized to the signal when all atoms were repumped to the state F=2.

\section{\label{appC}Frequency chain}

The 6.835 GHz frequency used to coherently manipulate the atomic spin is generated by a frequency chain based on a Spectra Dynamics DLR-100 system as a frequency reference. The
DLR-100 relies on an ultra-low noise 100 MHz quartz, locked at low frequency to the 10$^{\rm{th}}$ harmonic of a frequency doubled 5 MHz quartz to further improve the phase
noise. The 100 MHz signal is multiplied to 7 GHz and then mixed with a tunable synthesizer at 165 MHz to obtain the signal resonant with the transition between the $\left| \,
\downarrow \, \right\rangle$ and $\left| \, \uparrow \, \right\rangle$ state. The noise added by the microwave pulses to the measurement of $J_z$ is negligible for our weak
probe, and its control well below the atomic quantum projection noise level has already been shown in \cite{chen2012}. Frequency noise is added to the LO signal using a
frequency modulation port on the synthesizer, with a conversion factor set to 200 Hz/V$_{\rm{rms}}$. The noise signal for the demonstration of the clock using the PLL sequence
is generated with a signal generator, which produces white frequency noise with a spectral density of 2.7$\times$10$^{-2}$ Hz$^2$/Hz; this signal is low pass filtered at 1.85
kHz before being added to the LO. The result is a rms phase drift of 430 mrad over 10 ms for the LO.

\section{\label{appD}Feedback controller}

The atomic populations on the $\left| \, \downarrow \, \right\rangle$ and $\left| \, \uparrow \, \right\rangle$ states determine a differential phase shift of the probe
sidebands. As a consequence, the probe beam is modulated in amplitude, and the modulation signal is detected by a photodiode (1591NF, New Focus), amplified (two HMC716LP3E,
Hittite) and demodulated (ZX05-73C-C+, Minicircuits). The electronic integration of this signal during the 60 ns interrogation pulse produces a voltage proportional to the
average atomic population difference during the probing. Such a voltage is digitized by a 14 bit resolution analog-to-digital converter embedded in the microcontroller unit (MCU) used
to control the feedback loop (ADuC841, Analog Devices). In the experiments implementing the phase lock between the LO and the atomic superposition the MCU controls the phase
actuator, a 6 bits step phase shifter with a range of 2$\pi$ (RFPSHT0204N6, RF-Lambda). The total delay for the feedback is approximately 150 $\mu$s depending on the calculation
time of the MCU. When running the clock based on the PLL technique, the MCU acts as well on a frequency actuator, which is the frequency modulation input on the 165 MHz
synthesizer.

The feedback controller we propose for the clock exploiting the phase lock between the LO and the atomic superposition state consists of a cycle divided in three main steps: in the first one, successive correlated interrogations with probe interval T are realized on the same coherent atomic ensemble, and feedback on the LO phase is applied. The control law is
\begin{eqnarray}
\varphi_{\rm{LO}}^{(i)}	& =	& \varphi_{\rm{LO}}^{(i-1)} + \varphi_{\rm{FB}}^{(i)} \nonumber \\
			& =	& \varphi_{\rm{LO}}^{(i-1)} + g_{\varphi} \varphi^{(i)} \nonumber
\end{eqnarray}
where $\varphi^{(i)}$ is the estimated phase difference between the LO and the atoms at the i-\textit{th} cycle and $\varphi_{\rm{LO}}^{(i)}$ is the phase of the LO only. The values of $\varphi_{\rm{FB}}^{(i)} = g_{\varphi} \varphi^{(i)}$ are saved in the feedback controller and typically a gain $g_{\varphi}=-1$ is chosen.

The second step consists in the final phase readout: at the end of the N-\textit{th} interrogation interval of duration T, a destructive measurement $\varphi_{\rm{f}}$ of the phase is performed, with the highest possible precision. The saved phase shifts on the LO and $\varphi_{\rm{f}}$ are used to reconstruct the full phase drift between the LO and the atoms in the total interrogation time, equal to T$_{\rm{tot}}$=N$\times$T.

In the third step, one then performs feedback on the frequency as in a conventional atomic clock
\begin{eqnarray}
\omega_{\rm{LO}}^{(n)}	& =	& \omega_{\rm{LO}}^{(n-1)} + \omega_{\rm{FB}}^{(n)} \nonumber \\ 
				& =	& \omega_{\rm{LO}}^{(n-1)} + g_{\omega} \varphi_{\rm{tot}}^{(n)} / \rm{T}_{\rm{tot}}\nonumber
\end{eqnarray}
where
\begin{eqnarray}
\varphi_{\rm{tot}}^{(n)}	& =	& \varphi_{\rm{f}}^{(n)} - \sum_{i \in n-\rm{\textit{th} \; cycle}} \varphi_{\rm{FB}}^{(i)} \nonumber \\
				& =	& \varphi_{\rm{f}}^{(n)} - \sum_{i \in n-\rm{\textit{th} \: cycle}} g_{\varphi} \varphi^{(i)} \nonumber
\end{eqnarray}
and $n$ is the clock cycle. In addition, the phase shift set by the feedback controller on the LO is reset to zero
\begin{equation}
\varphi_{\rm{LO}}^{(n)} = \varphi_{\rm{LO}}^{(0)} = 0 \, \nonumber .
\end{equation}
The cycle then repeats. The important feature of the feedback controller is that the feedback actions on the LO oscillator phase during the interrogation time are saved. They
are then used with the output of the final precise measurement to determine the total phase drift. There is no drawback from the uncertainty of the weak measurements, since
any feedback errors are detected with the precise final measurement at the end. As already remarked, in the experimental demonstration of the clock operation, we adopted the
same probe for the intermediate and the final measurements.

\bibliographystyle{apsrev-title}

\begin{thebibliography}{41}
\bibitem[]{note1}
  \bibinfo{note}{We remark that an unambiguous determination of the LO phase drift beyond the inversion region could be obtained by tracking its evolution with periodic measurements of two orthogonal projections of the phase, at least when the Bloch vector is close to the borders of the inversion region. This solution does not require the phase lock protocol, but brings an increase in the number of interrogations of the atomic sample, and hence of the probe induced destructivity}.
\end{thebibliography}

\end{document}